\documentclass{aa}  

\usepackage{graphicx}
\usepackage{txfonts}
\usepackage{color}
\usepackage{natbib}
\begin{document} 

\title{
        Mass estimates for very cold (< 8 K) gas in molecular cloud cores
      }

\author{
J. Steinacker
\inst{1,2,3}
\and
H. Linz
\inst{3}
\and
H. Beuther
\inst{3}
\and
Th. Henning
\inst{3}
\and
A. Bacmann
\inst{1,2}
       }

\institute{
Univ. Grenoble Alpes, IPAG, F-38000 Grenoble, France\\
\email{stein@mpia.de}
\and
CNRS, IPAG, F-38000 Grenoble, France
\and
Max-Planck-Institut f\"ur Astronomie,
K\"onigstuhl 17, D-69117 Heidelberg, Germany
          }

\date{Received; accepted}

\abstract
{
The mass of prestellar cores is an essential ingredient to
understand the onset of star formation in the core. The low level of emission
from cold dust may keep parts of this dust hidden from observation.
}
{
We aim to determine the fraction of core mass in the temperature range 
$<$ 8 K
that can be expected for typical low- and high-mass star formation regions.
}
{
We calculated the dust temperature within 
standard spherically symmetric 
prestellar cores for a grid of density power laws in the outer core
regions, core masses, and variations in the external multicomponent
radiation field.
We assume the dust is composed of amorphous silicate and carbon 
and we discuss variations of its optical properties. 
As a measure for the distribution of cores and clumps, we used core mass 
functions derived for various environments.
In view of the high densities in very cold central regions, dust and gas temperatures
are assumed to be equal.
}
{
We find that the fraction of mass with temperatures $<$ 8 K 
in typical low- and high-mass cores is $<$ 20\%.
It is possible to obtain higher fractions of very cold gas by placing 
intermediate- or high-mass cores in a typical low-mass star formation environment.
We show that the mass uncertainty arising from far-infrared (FIR) to mm modeling of very cold dust 
emission is smaller than the mass uncertainty owing to the unknown dust opacities.
}
{
Under typical star formation conditions, dust with temperatures $<$ 8 K 
covers a small mass fraction in molecular cloud cores, but may play a more 
important role for special cases.
The major unknown in determining the total core mass from thermal dust emission
is the uncertainty
in the dust opacity, not in the underestimated very cold dust mass.
}

\keywords{
        Radiative transfer --
        Radiation mechanisms: thermal --
Stars: formation --
ISM: clouds
         }

\maketitle
 
\section{Introduction}
While molecular cloud cores have been identified as the site where the
star formation starts, which physical processes
initiate and control the onset is still a matter of debate.
To assess the impact of effects such as gravitational instability or shielding
from the outer radiation field, the spatial mass distribution of the core is one
of the key ingredients to progress to a paradigm.
Since molecules tend to freeze out onto the dust grains in the shielded cold core center 
with temperature around 10 K, thermal emission from dust is the main
tool to determine the core structure and its mass.

Accumulated for an entire star formation region, the numbers of cores in a certain
mass range can aid the understanding of the overall efficiency of the star formation process
when compared to the initial mass function of stars
\citep[e.g.,][]{1998A&A...336..150M, 2007A&A...462L..17A}.
\citet{2010A&A...518L.106K}, for example, have determined such a core mass function (CMF)
for Aquila based
on single-temperature modified blackbody fits of the dust emission maps.
Also, CMFs were derived
 for cores in a massive star formation environment 
\citep[e.g., for Orion A North;][]{2007MNRAS.374.1413N}.
Computationally, more costly improvements 
to the single-temperature approach consider a spherical or
elliptical core model with a radial dust temperature profile
\citep{
2001ApJ...557..193E,
2001A&A...376..650Z,
2005ApJ...632..982S,
2012A&A...547A..11N,
2013A&A...551A..98L, 
2014A&A...562A.138R, 
Stei16}.
Already in the early work by \citet{2001ApJ...557..193E} and \citet{2001A&A...376..650Z}, 
the dust temperature distribution of prestellar cores is calculated
self-consistently by illuminating the core with an external radiation field.
\citet{2001A&A...376..650Z} proposed a multicomponent model of the
interstellar radiation field (ISRF) based on the data given in
\citet{1994ASPC...58..355B}.
In their discussion they stressed that the model likely underestimates both the 
far-infrared (FIR) and mid-infrared (MIR)
field caused by local sources such as thermal emission from big
dust grains or small stochastically-heated grains (SHG).
Without this component, Zucconi et al. derived temperatures $<$ 8 K for core regions 
with  visual extinction >30 mag. 
They concluded that the emission seen in FIR maps may be dominated by the
contribution from the low-density exterior of the core.

The argument was repeated by
\citet{2015A&A...574L...5P} when 
comparing FIR and submillimeter (sub-mm) emission across the core L183 to the 8 $\mu$m
absorption map from Spitzer data. Fitting modified blackbody functions toward three 
different positions, these authors found that the 
contribution of a cold dust component with temperatures $<$ 10 K were not
well constrained.
They expressed their belief that the two cores L183 and L1689B are clear 
illustrations of a general problem of detecting the very cold gas mass reservoir in
prestellar cores. They concluded that mass measurements based on dust emission alone
might miss a large fraction of the core (30\% to 70\% for 
L183 and L1689B).

However, 
a detailed modeling of the filament-embedded prestellar core L1689B 
\citep{Stei16}, based on Herschel PACS/SPIRE and JCMT/SCUBA-2 maps revealed
that the gas mass content at temperatures $<$ 8 K is less than
10\%, taking the full ambiguity of low emissivity cold dust into account.
In the same work, a synergetic radiative transfer approach also allowed the investigators
to derive the local ISRF of L1689B with warm and cold dust components that were
a factor of 4 higher than the standard ISRF proposed by 
\citet{1994ASPC...58..355B} (see also Fig.~\ref{ISRFvar}).
This might be the main reason for \citet{2001ApJ...557..193E} to find the temperature
in L1689B dropping below 8 K in the inner few 1000 au; these authors assumed, for their
best fit, an outer field that is a factor of up to 8 lower than the field derived
in \citet{Stei16} based on the Herschel/JCMT maps (their Fig.~5). Differences in
the assumed core and dust properties may contribute to the deviation as well.
Nevertheless, even with their weak outer illumination, \citet{2001ApJ...557..193E}
find only a gas mass of about 5\% at a temperature below 8 K.

In this work, we investigate what gas mass fraction is expected
to host dust at temperatures $<$ 8 K for a typical prestellar core. 
While we derive dust temperatures, in regions where
the dust is at very low temperatures of $<$ 8 K the densities are high enough
that gas and dust temperatures are the same. In this way, we estimate
the very cold gas masses that contain very cold dust using the assumed
gas-to-dust mass ratio (120 is used in this work). 
The choice of the actual value of 8 K as a border for very cold dust
is somewhat arbitrary in view of the continuous 
decrease of the emission efficiency with temperature. We therefore
follow earlier work that defined very cold dust to have temperatures 
up to 8 K to make the comparison easier
\citep{2015A&A...574L...5P}.

In Sect.~\ref{222}, we define the prestellar cores, ISRF models, and dust properties, and we describe the numerical approach.
Sect.~\ref{333} contains the obtained radial temperature profiles and 
mass fractions of gas with dust temperatures $<$ 8 K. 
We discuss the findings and present our conclusions in Sect.~\ref{444}.

\section{Models description and numerical scheme}\label{222}
For this assessment of typical prestellar cores,
we use a simple spherically symmetric structure model
although the geometry is often more complex
\citep[see, e.g,][]{2005A&A...434..167S}. 
Deviations are discussed in the Sect.~\ref{444}.
As the H$_2$ number density
profile, we choose the typical flat-to-power-law dependence of the form
$n(r)=n_0/[1+(r/r_0)^p]$ cut off at $r=R_c$ 
\citep[see, e.g.,][]{2000A&A...361..555B}.
The normalization $n_0$ is adjusted so that
an integral over the gas in the core volume matches the chosen core mass $M_c$.
The exponent $p$ is varied between 1 and 3, and the kink radius $r_0$, where
the flat profile changes to the power-law profile, is chosen as $R_c/4$.
Because of the observed increase of the average core radius with mass, we
parameterize the outer radius as $R_c=[1+2log_{10}(M_c/M_0)]\times 10^4$ au
(based on a log interpolation between cores having outer radii 
of 10$^4$ au (7$\times$10$^4$ au) for a gas mass of 
0.1 M$_\odot$ (100 M$_\odot$), respectively. The
considered mass range is $M_0$=0.1 M$_\odot$ to $M_1$=100 M$_\odot$.
The outer core radius is not well defined 
\citep{2007A&A...476.1243M,2014prpl.conf..149T}, 
but we chose the high value
7$\times$10$^4$ au to consider the shielding by the outer clump as well.

Since the dust opacities are needed from the UV to the mm,
we rely on the silicate and carbon data by 
\citet{1984ApJ...285...89D}
but assume an MRN size-distribution 
\citep{1977ApJ...217..425M}
up to sizes of 0.8 $\mu$m to account for
additional scattering often found in prestellar cores and interpreted
to arise from coagulated grains 
\citep{2010A&A...511A...9S}. 
The resulting opacity is a about a factor of 2.5 lower than OH5 
\citep{1994A&A...291..943O} at 1 mm.
The gas-to-dust mass ratio values used in the literature vary from 
        100 \citep{2014A&A...562A.138R}
        over 133 \citep{2011A&A...525A.103C}
        to 150 \citep{2011piim.book.....D}. In this work we used 120.
Thin ice mantles are expected to increase the opacities
for the FIR/sub-mm wavelength range, however, it has to be kept in mind that the nature of 
the dust grains in prestellar cores is still unknown and even 
the various dust model opacities vary by a factor of 3.

\begin{figure}
\vbox{
\includegraphics[width=9cm]{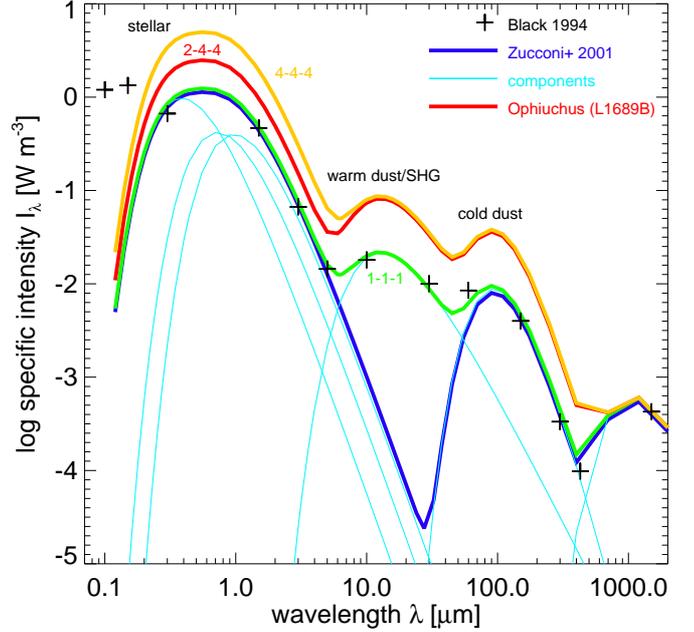}
     }
\caption{
     Interstellar radiation field model featuring a stellar (I), 
     warm dust/SHG (II), cold dust (III), and cosmic background component
     after \citet{2001A&A...376..650Z}. The total field  is shown in blue with the
     different components as magenta lines. A warm dust component was added
     to fit the data by \citet{1994ASPC...58..355B} (plus signs).
     The green (orange) curve shows our low(high)-mass ISRF model,
     respectively (notations of the component factors: I-II-III).
     The red curve shows the ISRF determined for L1689B in Ophiuchus containing
     B stars, but not near the core \citep{Stei16}.
        }
     \label{ISRFvar}
     \end{figure}
The ISRF for low- and high-mass star formation regions is expected to show strong local
variation due to nearby stellar, warm dust, or stochastically heated grain
contributions, which may at shorter wavelengths be modified because of 
extinction and scattering.
As an approximative model we used 
the field by \citet{1994ASPC...58..355B}
that was modeled with single components as described in \citet{2001A&A...376..650Z}.
To take into account the local variations, we can multiply
each of the components for stellar radiation (I), warm dust (II), and 
cold dust (III) by a multiplicative factor (see Fig.~\ref{ISRFvar}) (notation for the component factors: I-II-III).
Here, we discuss two cases: a typical ISRF for a low-mass star formation region (green, 1-1-1)
and for a high-mass star formation region (orange, 4-4-4). 
As an example, we add the local field of L1689B
\citep[][red, 2-4-4]{Stei16} indicating that PDRs
strongly enhance the MIR/FIR field, but that the core is far enough from
the B2 stars in that region to see just a moderate increase in the stellar
component. 
We note that the (4-4-4) field is likely only a lower limit on the radiation
field in massive star formation regions
\citep{2006A&A...449..609J}, as these amplification factors are almost 
reached already, for example, in the side-complex L1689 of the Ophiuchus cloud
\citep{Stei16},
compared to \citet{1999A&A...344..342L} who give a single amplification factor of 10 for the main cloud part L1688.
But since we want to derive upper limits for the mass of very cold gas this
is a conservative approach.

To perform a large grid of core models with different mass and outer density
power-law exponent,
we use a one-dimensional (1D) ray-tracing code that transports the external radiation to all 
core cells, and then also derives the correct dust temperature in
the optically thick case by iterated cell-cell illumination. Modifications
due to a three-dimensional (3D) structure are discussed in the conclusions.
We include scattering in the extinction but we do not follow scattered
radiation. The radiation field in the cold region ($<$ 10 K) of the cores is
dominated by radiation from a wavelength range where scattering is no 
longer important and does not contribute to the heating. Regions where 
scattered radiation increases the temperature may indicate lower temperatures when
scattered radiation is neglected. In the case of self-absorption (for
higher mass cores) this may lead to a slight underestimation of the 
temperature in the outer core region that hardly affects our derived upper
mass limits for very cold gas.

\section{Radial temperature profiles and mass fractions}\label{333}
We first consider a low-mass star formation region characterized by
a (1-1-1) ISRF.
\begin{figure}
\vbox{
\includegraphics[width=9cm]{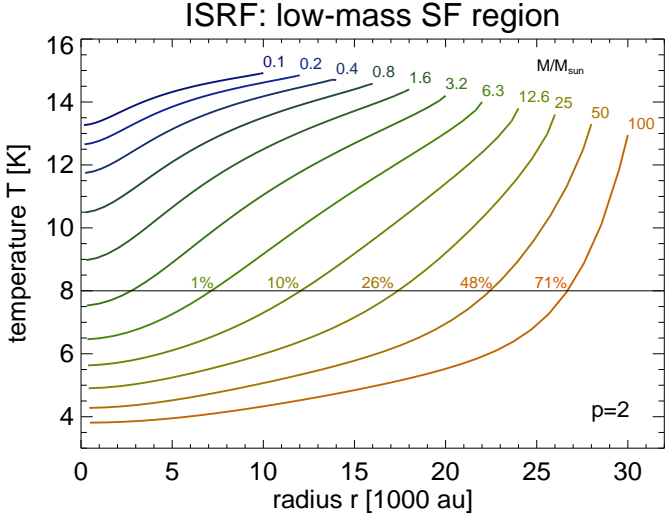}
     }
\caption{
Dust temperature profiles of cores with different total masses in a 
typical low-mass ISRF (1-1-1). The density power-law index was -2 (p=2)
in the outer core regions. The horizontal line indicates the dust temperature 
limit of 8 K considered in this work; the percentage numbers indicate the
fraction of mass with gas below this temperature.
        }
\label{Tlow}
\end{figure}
Fig.~\ref{Tlow} shows the dust temperature profiles of cores for various
masses with a power-law exponent -2 (e.g., in the isothermal sphere
and Bonnor-Ebert sphere model). A horizontal line at 8 K indicates at which
radii the profiles enter the region of very cold dust and gas,
as the gas has the same temperature as the dust. 
The percentage numbers give the gas mass for which the temperature
is $<$ 8 K.
The plot shows that cores with a total mass of 50 $M_\odot$
in a low-mass star formation region
can have about half of their gas mass at T $<$ 8 K.
Temperatures below 5 K are likely prevented by the
action of cosmic ray heating \citep[][Sect.~5]{2001ApJ...557..193E}.

\begin{figure}
\vbox{
\includegraphics[width=9cm]{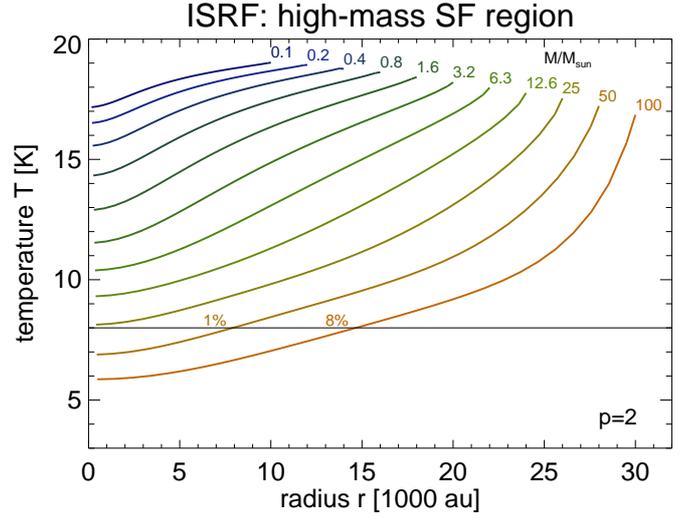}
     }
\caption{
Same as Fig.~\ref{Tlow} for a high-mass (4-4-4)-ISRF.
        }
\label{Thigh}
\end{figure}
Exposed to the (4-4-4) ISRF in a high-mass star formation region, 
the dust temperature profiles rise compared to the low-mass case
(Fig.~\ref{Thigh}).
As a result, even the high-mass cores show little mass at T $<$ 8 K.

Since the temperature depends on the column density, the gradient
of the density profile affects the findings. To explore this
effect, we ran a grid of core masses and density power index p.
For each core, we determined the fraction of the core gas mass
with dust at temperature $<$ 8K.
\begin{figure}
\vbox{
\includegraphics[width=9cm]{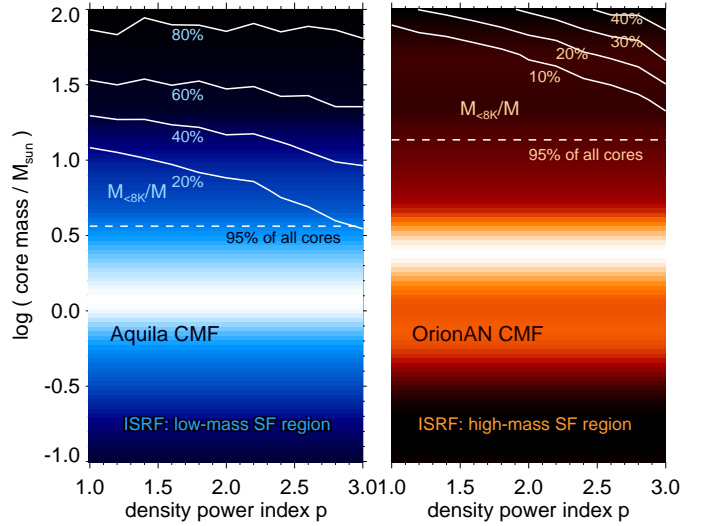}
     }
\caption{
Contours: mass fraction of gas with temperatures $<$ 8 K as a 
function of total core mass, and density power-index p for a low-mass star formation
(1-1-1) ISRF (left) and a high-mass star formation (4-4-4) ISRF (right), respectively.
Color-coded background: 
for each $p$, the core mass functions of Aquila (left) and
Orion A North (right) are shown as a function of the core mass.
Since the actual $p$ dependence is unknown, we assume that all cores have the same $p$.
The dashed line indicates the mass limit below which
95\% of all cores are located.
        }
\label{Mp}
\end{figure}
The result is shown as contour lines in Fig.~\ref{Mp} for the low- and high-mass 
star formation case at left
and right, respectively. The labels indicate the percentage of very cold gas.
The general trend is that core with steeper density gradients show higher
mass percentages of very cold gas. 

To relate this limit to typical core mass distributions in
the two cases, we
added color-coded CMFs as background for Aquila (Orion A North) in the
left (right) contour plot, respectively. 
Since the density power index $p$ has not been measured for all cores and
likely also varies within each core,
we show the color-coded CMFs along the mass axis
assuming that all cores have the same $p$.
The real distribution is likely a more concentric pattern around
the often observed p=2.
The plot shows that even when assuming a steep gradient of $p=3$ for all
cores in Aquila, the mass fraction of very cold gas hardly gets above 20\% for
95\% of all cores (dashed line). For the cores in the high-mass region,
even the 10\% limit is not reached at the extreme value of $p=3$.

\section{Discussion and conclusions}\label{444}
The variations in the local ISRF of the cores are expected to have a 
larger range than the two cases (1-1-1) and (4-4-4) discussed here.
It is clear from the presented temperature profiles that any additional
local source, such as a nearby star or PDR, further reduces the amount of 
very cold gas.

On the other hand, star-forming regions host also large amounts of
matter shielding the core from the
incoming local ISRF. This {\em core-external extinction} acts on top
of the extinction inside the core. 
\citet{2001A&A...376..650Z} explicitly consider an optical extinction
of 1-2 mag due to large-scale diffuse cloud dust suppressing the UV part
of the \citet{1994ASPC...58..355B} ISRF composite. On smaller scales,
filaments hosting the cores create
additional increased extinction regions in the direction of the filament axis.
Filaments are optically thin in low-mass star formation regions so
that the MIR and FIR components that are responsible for the heating
of the core center are not affected. The filament can be more opaque for high-mass star formation
regions.
\citet{2013A&A...557A.120K} find for the "snake" infra-red dark cloud G11.11-0.12
that about 2$\times$10$^4$ M$_\odot$ is located in the 30 pc long filament,
so that a range of 1-8 mag of extinction seems possible before the radiation
field can start heating the cores inside.
The actual value depends on the local clumpiness in a certain direction.
In turn the local MIR component in particular is also be enhanced in high-mass star formation regions.
\citet{Lippok}, for example, determined the temperature profiles of six prestellar cores
with masses ranging from 2.6 to 14 M$_\odot$. Since they are located in very
different environments their local fields are found to vary by application factors
from 1 to 3.
The derived temperature profiles show that none of the cores has a noteworthy mass
fraction of gas at temperatures $<$ 8 K.

Nevertheless,
there is the possibility that a low-mass core is surrounded by
a more massive, but extended envelope. In this case, the shielding and
thus the central temperature would be that of a more massive core.
Correspondingly, the central region may host a substantial fraction of very
cold gas (see Fig.~\ref{Tlow} for masses > 20 M$_\odot$).
L183 is such a case with a complex gas structure of about 80 M$_\odot$
\citep{2004A&A...417..605P}
and this object is exposed to radiation
that is expected to be  only slightly higher than the standard ISRF (because of its elevation of 36$^o$ above the Galactic plane, but no proximity to PDRs or
early-type stars, so likely a reduced warm dust component).

However, the actual 3D structure of the considered core strongly influences this. If
there are "holes" allowing radiation to penetrate deeper, the temperatures
remains higher \citep[for a discussion of clumpy shells 
see, e.g.,][]{2005prpl.conf.8255I}.

Another 3D structure effect concerns a possible fragmentation in the central 
        few thousand au.
Compared to our simple 1D model density this places more gas in better shielded
regions and thus the mass in cold gas could be higher. It has to be kept in mind, 
however, that fragmentation in the central region does not
enclose a large gas mass compared to the total core mass. 
Moreover, the clumpiness also allows the radiation field to penetrate deeper, which
partially counteracts the stronger shielding. A precise prediction is
only possible when making assumptions about the clumpiness, which is beyond the
scope of this letter.

Concerning the assumed opacity, it could be argued that a higher opacity
would increase extinction and therefore improve shielding and thus favor 
colder central temperatures.
We performed tests to find out how much the obtained mass fractions of very cold dust
vary with an increase in the dust opacity.
We find no strong dependence on the opacity change by factors of a few.
The core center is mainly heated
by external FIR/mm radiation and then the balance equation from which T is
derived contains the opacity 
on both sides.
For massive cores, an additional radiation source is the outer core region,
where most of the UV-to-MIR radiation is converted to FIR radiation.
Varying the opacity changes the temperature structure of this outer part,
but only mildly changes its FIR radiation output.

In conclusion, for the bulk of cores in the low- and high-mass case we find
that with reasonable assumptions about the incident local radiation field
the mass fraction with dust at temperatures below 8 K exceeds
20\% only for the high-mass tail of the distribution representing 5\% of all
cores. 
While this dust is more difficult to detect in FIR/sub-mm data,
it comprises a small uncertainty in the total core mass determination compared
to the uncertainty of the dust properties resulting in factors of a few.
Also, the uncertainty from very cold gas is 
smaller than the uncertainties in the mass due to incorrect inclusion
of surrounding filament gas, which can yield factors of 2
\citep[see, e.g.,][]{Stei16}.

\begin{acknowledgements}
We thank the referee for very constructive comments that helped improve
the paper.
\end{acknowledgements}

\bibliographystyle{aa} 
\bibliography{lowt} 

\begin{thebibliography}{29}
\expandafter\ifx\csname natexlab\endcsname\relax\def\natexlab#1{#1}\fi

\bibitem[{{Alves} {et~al.}(2007){Alves}, {Lombardi}, \&
  {Lada}}]{2007A&A...462L..17A}
{Alves}, J., {Lombardi}, M., \& {Lada}, C.~J. 2007, \aap, 462, L17

\bibitem[{{Bacmann} {et~al.}(2000){Bacmann}, {Andr{\'e}}, {Puget}, {Abergel},
  {Bontemps}, \& {Ward-Thompson}}]{2000A&A...361..555B}
{Bacmann}, A., {Andr{\'e}}, P., {Puget}, J.-L., {et~al.} 2000, \aap, 361, 555

\bibitem[{{Black}(1994)}]{1994ASPC...58..355B}
{Black}, J.~H. 1994, in Astronomical Society of the Pacific Conference Series,
  Vol.~58, The First Symposium on the Infrared Cirrus and Diffuse Interstellar
  Clouds, ed. R.~M. {Cutri} \& W.~B. {Latter}, 355

\bibitem[{{Compi{\`e}gne} {et~al.}(2011){Compi{\`e}gne}, {Verstraete}, {Jones},
  {Bernard}, {Boulanger}, {Flagey}, {Le Bourlot}, {Paradis}, \&
  {Ysard}}]{2011A&A...525A.103C}
{Compi{\`e}gne}, M., {Verstraete}, L., {Jones}, A., {et~al.} 2011, \aap, 525,
  A103

\bibitem[{{Draine}(2011)}]{2011piim.book.....D}
{Draine}, B.~T. 2011, {Physics of the Interstellar and Intergalactic Medium}

\bibitem[{{Draine} \& {Lee}(1984)}]{1984ApJ...285...89D}
{Draine}, B.~T. \& {Lee}, H.~M. 1984, \apj, 285, 89

\bibitem[{{Evans} {et~al.}(2001){Evans}, {Rawlings}, {Shirley}, \&
  {Mundy}}]{2001ApJ...557..193E}
{Evans}, II, N.~J., {Rawlings}, J.~M.~C., {Shirley}, Y.~L., \& {Mundy}, L.~G.
  2001, \apj, 557, 193

\bibitem[{{Indebetouw} {et~al.}(2005){Indebetouw}, {Whitney}, {Johnson}, \&
  {Wood}}]{2005prpl.conf.8255I}
{Indebetouw}, R., {Whitney}, B.~A., {Johnson}, K.~E., \& {Wood}, K. 2005, in
  Protostars and Planets V Posters, Vol. 1286, 8255

\bibitem[{{J{\o}rgensen} {et~al.}(2006){J{\o}rgensen}, {Johnstone}, {van
  Dishoeck}, \& {Doty}}]{2006A&A...449..609J}
{J{\o}rgensen}, J.~K., {Johnstone}, D., {van Dishoeck}, E.~F., \& {Doty}, S.~D.
  2006, \aap, 449, 609

\bibitem[{{Kainulainen} {et~al.}(2013){Kainulainen}, {Ragan}, {Henning}, \&
  {Stutz}}]{2013A&A...557A.120K}
{Kainulainen}, J., {Ragan}, S.~E., {Henning}, T., \& {Stutz}, A. 2013, \aap,
  557, A120

\bibitem[{{K{\"o}nyves} {et~al.}(2010){K{\"o}nyves}, {Andr{\'e}},
  {Men'shchikov}, {Schneider}, {Arzoumanian}, {Bontemps}, {Attard}, {Motte},
  {Didelon}, {Maury}, {Abergel}, {Ali}, {Baluteau}, {Bernard}, {Cambr{\'e}sy},
  {Cox}, {di Francesco}, {di Giorgio}, {Griffin}, {Hargrave}, {Huang}, {Kirk},
  {Li}, {Martin}, {Minier}, {Molinari}, {Olofsson}, {Pezzuto}, {Russeil},
  {Roussel}, {Saraceno}, {Sauvage}, {Sibthorpe}, {Spinoglio}, {Testi},
  {Ward-Thompson}, {White}, {Wilson}, {Woodcraft}, \&
  {Zavagno}}]{2010A&A...518L.106K}
{K{\"o}nyves}, V., {Andr{\'e}}, P., {Men'shchikov}, A., {et~al.} 2010, \aap,
  518, L106

\bibitem[{{Launhardt} {et~al.}(2013){Launhardt}, {Stutz}, {Schmiedeke},
  {Henning}, {Krause}, {Balog}, {Beuther}, {Birkmann}, {Hennemann},
  {Kainulainen}, {Khanzadyan}, {Linz}, {Lippok}, {Nielbock}, {Pitann}, {Ragan},
  {Risacher}, {Schmalzl}, {Shirley}, {Stecklum}, {Steinacker}, \&
  {Tackenberg}}]{2013A&A...551A..98L}
{Launhardt}, R., {Stutz}, A.~M., {Schmiedeke}, A., {et~al.} 2013, \aap, 551,
  A98

\bibitem[{{Lippok} {et~al.}(2016){Lippok}, {Launhardt}, {Henning}, {Beuther},
  {Kainulainen}, {Krause}, {Linz}, {Nielbock}, {Ragan}, {Robitaille},
  {Sadavoy}, \& {Schmiedeke}}]{Lippok}
{Lippok}, N., {Launhardt}, R., {Henning}, T., {et~al.} 2016, ArXiv e-prints
  1606.04318

\bibitem[{{Liseau} {et~al.}(1999){Liseau}, {White}, {Larsson}, {Sidher},
  {Olofsson}, {Kaas}, {Nordh}, {Caux}, {Lorenzetti}, {Molinari}, {Nisini}, \&
  {Sibille}}]{1999A&A...344..342L}
{Liseau}, R., {White}, G.~J., {Larsson}, B., {et~al.} 1999, \aap, 344, 342

\bibitem[{{Mathis} {et~al.}(1977){Mathis}, {Rumpl}, \&
  {Nordsieck}}]{1977ApJ...217..425M}
{Mathis}, J.~S., {Rumpl}, W., \& {Nordsieck}, K.~H. 1977, \apj, 217, 425

\bibitem[{{Motte} {et~al.}(1998){Motte}, {Andre}, \&
  {Neri}}]{1998A&A...336..150M}
{Motte}, F., {Andre}, P., \& {Neri}, R. 1998, \aap, 336, 150

\bibitem[{{Motte} {et~al.}(2007){Motte}, {Bontemps}, {Schilke}, {Schneider},
  {Menten}, \& {Brogui{\`e}re}}]{2007A&A...476.1243M}
{Motte}, F., {Bontemps}, S., {Schilke}, P., {et~al.} 2007, \aap, 476, 1243

\bibitem[{{Nielbock} {et~al.}(2012){Nielbock}, {Launhardt}, {Steinacker},
  {Stutz}, {Balog}, {Beuther}, {Bouwman}, {Henning}, {Hily-Blant},
  {Kainulainen}, {Krause}, {Linz}, {Lippok}, {Ragan}, {Risacher}, \&
  {Schmiedeke}}]{2012A&A...547A..11N}
{Nielbock}, M., {Launhardt}, R., {Steinacker}, J., {et~al.} 2012, \aap, 547,
  A11

\bibitem[{{Nutter} \& {Ward-Thompson}(2007)}]{2007MNRAS.374.1413N}
{Nutter}, D. \& {Ward-Thompson}, D. 2007, \mnras, 374, 1413

\bibitem[{{Ossenkopf} \& {Henning}(1994)}]{1994A&A...291..943O}
{Ossenkopf}, V. \& {Henning}, T. 1994, \aap, 291, 943

\bibitem[{{Pagani} {et~al.}(2004){Pagani}, {Bacmann}, {Motte}, {Cambr{\'e}sy},
  {Fich}, {Lagache}, {Miville-Desch{\^e}nes}, {Pardo}, \&
  {Apponi}}]{2004A&A...417..605P}
{Pagani}, L., {Bacmann}, A., {Motte}, F., {et~al.} 2004, \aap, 417, 605

\bibitem[{{Pagani} {et~al.}(2015){Pagani}, {Lef{\`e}vre}, {Juvela}, {Pelkonen},
  \& {Schuller}}]{2015A&A...574L...5P}
{Pagani}, L., {Lef{\`e}vre}, C., {Juvela}, M., {Pelkonen}, V.-M., \&
  {Schuller}, F. 2015, \aap, 574, L5

\bibitem[{{Roy} {et~al.}(2014){Roy}, {Andr{\'e}}, {Palmeirim}, {Attard},
  {K{\"o}nyves}, {Schneider}, {Peretto}, {Men'shchikov}, {Ward-Thompson},
  {Kirk}, {Griffin}, {Marsh}, {Abergel}, {Arzoumanian}, {Benedettini}, {Hill},
  {Motte}, {Nguyen Luong}, {Pezzuto}, {Rivera-Ingraham}, {Roussel}, {Rygl},
  {Spinoglio}, {Stamatellos}, \& {White}}]{2014A&A...562A.138R}
{Roy}, A., {Andr{\'e}}, P., {Palmeirim}, P., {et~al.} 2014, \aap, 562, A138

\bibitem[{{Shirley} {et~al.}(2005){Shirley}, {Nordhaus}, {Grcevich}, {Evans},
  {Rawlings}, \& {Tatematsu}}]{2005ApJ...632..982S}
{Shirley}, Y.~L., {Nordhaus}, M.~K., {Grcevich}, J.~M., {et~al.} 2005, \apj,
  632, 982

\bibitem[{{Steinacker} {et~al.}(2016){Steinacker}, {Bacmann}, {Henning}, \&
  {Heigl}}]{Stei16}
{Steinacker}, J., {Bacmann}, A., {Henning}, T., \& {Heigl}, S. 2016, ArXiv
  e-prints 1606.07942

\bibitem[{{Steinacker} {et~al.}(2005){Steinacker}, {Bacmann}, {Henning},
  {Klessen}, \& {Stickel}}]{2005A&A...434..167S}
{Steinacker}, J., {Bacmann}, A., {Henning}, T., {Klessen}, R., \& {Stickel}, M.
  2005, \aap, 434, 167

\bibitem[{{Steinacker} {et~al.}(2010){Steinacker}, {Pagani}, {Bacmann}, \&
  {Guieu}}]{2010A&A...511A...9S}
{Steinacker}, J., {Pagani}, L., {Bacmann}, A., \& {Guieu}, S. 2010, \aap, 511,
  A9

\bibitem[{{Tan} {et~al.}(2014){Tan}, {Beltr{\'a}n}, {Caselli}, {Fontani},
  {Fuente}, {Krumholz}, {McKee}, \& {Stolte}}]{2014prpl.conf..149T}
{Tan}, J.~C., {Beltr{\'a}n}, M.~T., {Caselli}, P., {et~al.} 2014, Protostars
  and Planets VI, 149

\bibitem[{{Zucconi} {et~al.}(2001){Zucconi}, {Walmsley}, \&
  {Galli}}]{2001A&A...376..650Z}
{Zucconi}, A., {Walmsley}, C.~M., \& {Galli}, D. 2001, \aap, 376, 650

\end{thebibliography}

\end{document}